# Millimeter Wave Channel Measurements in a Railway Depot

Berna Bulut, Thomas Barratt, Di Kong, Jue Cao, Alberto Loaiza Freire,
Simon Armour, Mark Beach

Communication Systems & Networks,
University of Bristol, Bristol, United Kingdom

Email: berna.bulut@bristol.ac.uk

*Abstract*—Millimeter wave (mmWave) communication is a key enabling technology with the potential to deliver high capacity, high peak data rate communications for future railway services. Knowledge of the radio characteristics is of paramount importance for the successful deployment of such systems. In this paper mmWave channel measurements are reported for a railway environment using a wideband channel sounder operating at 60GHz. Highly directional antennas are deployed at both ends of the link. Data is reported for path loss, root mean square (RMS) delay spread and K-factor. Static and mobile measurements are considered. Analysis shows that the signal strength is strongly dependent (up to 25dB) on the azimuth orientation of the directional transmit and receive antennas. A path loss exponent of $n = 2.04$ was extracted from the Line-of-Sight measurements with optimally aligned antennas. RMS delay spreads ranged from 1ns to 22ns depending on antenna alignment. 50% of the measured K-factors were found to be less than 6dB. We conclude this is the result of ground reflections in the vertical Tx-Rx plane.

## I. INTRODUCTION

The provision of broadband Wi-Fi services is of growing importance to train operators. These can be used to enhance rail passenger experience, satisfaction and productivity. Providing Wi-Fi services inside a train is relatively simple, however the bottleneck arises from capacity limitations in the fronthaul link between the train and trackside infrastructure. The use of millimetre wave (mmWave) spectrum (24-300GHz) enables multi-gigabit-per-second communications to and from high-speed trains (HST). The IEEE 802.11ad standard [1] has already generated considerable interest in the use of 60GHz unlicensed spectrum [2]. The use of mmWave spectrum is also expected as part of the 5G New Radio (NR) standard [3].

The high transmission frequencies used in mmWave systems impose stringent limitations on the propagation range and overall system performance. Accurate knowledge of the electromagnetic propagation characteristics needs to be taken into account when designing and deploying future mmWave communication systems. Previous measurements have been reported for railway environments in the sub-6GH bands [4]. However, to the best of the authors' knowledge, no mmWave measurement campaigns have been reported at 60 GHz for railway environments. To address this gap, this paper presents a number of rail based mmWave measurements. In [5], traditional city-based measurements were reported at 23GHz and 73GHz. Based on this data statistical channel models were presented. In [6], measurements at 60GHz were performed for cellular peer-to-peer communications in an outdoor campus. However, limited information was provided on the statistical properties of the channel. Here we perform measurements at 60GHz in the St Philip's Marsh train depot in Bristol (UK). Our measurements were taken using highly directional antennas and are representative of a train-to-track communications link. This captured data is used to determine key mmWave channel parameters such as path loss, angle-of-arrival (AoA), Root Mean Square (RMS) delay spread and Rician K-factor. The work presented here is the first to provide outdoor multipath characteristics for wideband (1GHz) mmWave channels in a railway environment.

The remainder of this paper is organized as follows. Section II explains the details of our experimental setup. Measurement results and analysis is provided in Section III with conclusions presented in Section IV.

## II. EXPERIMENTAL CONFIGURATION

This section describes the configuration of the mmWave channel measurements, including an overview of the hardware and locations used.

### A. Hardware Configuration

A 1GHz wide baseband channel sounding signal was generated using the Keysight M9099 Waveform Creator software and used to configure a Keysight M8190A arbitrary waveform generator (ARB). This was then used to drive the I and Q ports on a Sivers IMA up-converter to generate an RF signal centred at 60GHz. The signal was then connected to a co-polarised circular horn antenna [7]. At the receiver, a circular horn antenna was connected to an orthomode transducer. Two Sivers IMA down-converters were used to receive both horizontal and vertical polarisations. The Sivers IMA devices down convert the 60GHz signal to an IQ IF signal. This is then captured and processed using a high performance digital oscilloscope (DSO) (a Keysight MSOS804A). The transmitter (Tx) and receiver (Rx) configurations are shown in Fig. 1 and Fig. 2 respectively. The transmitter and receiver employ 25dBi standard gain circular horn antennas with a 6.7° half-power beamwidth.

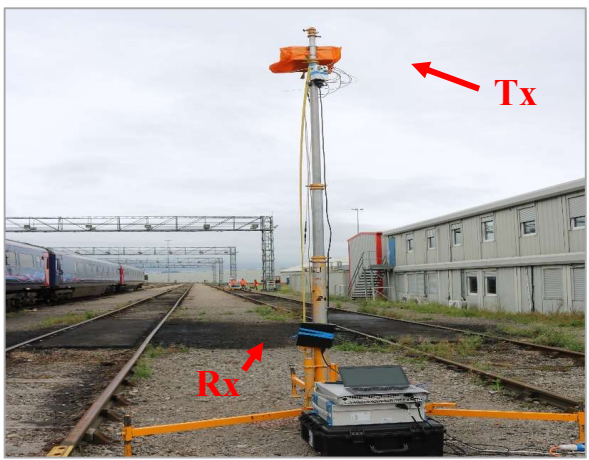

Fig. 1. 60 GHz transmitter setup using Keysight M8190 AWG and SiversIMA, and the measurement scenario.

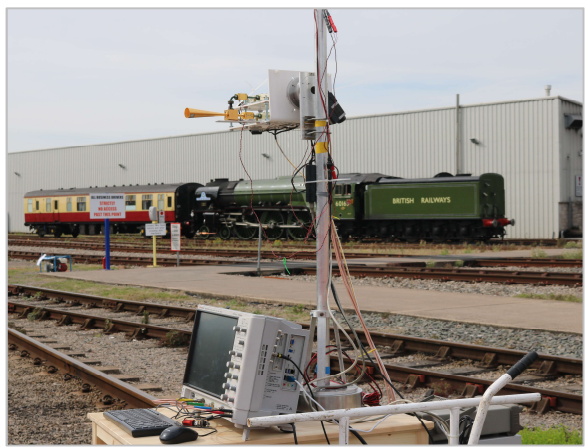

Fig. 2. Receiver architecture using orthomode transducer, Keysight MSOS804A 20 GHz oscilloscope and 89600 VSA software interface.

The Keysight 89600 VSA & Waveform Creator channel sounding function operates by repeatedly transmitting a single carrier signal bearing a modulated waveform. The waveform has excellent auto-correlation properties, and a low peak-to-average power ratio. The bandwidth and duration of the modulating waveform can be varied to suit the channel measurement required. Spectrum shaping can be applied to reduce out of band interference when transmitting the signal in a live environment.

*B. Measurment Scenario*

The channel sounding equipment described in Section II-A was used to measure the 60GHz outdoor propagation channel in a railway environment. The Tx was placed at the trackside as depicted in Fig. 1. The Tx antenna was mounted at a height of 4m on a pole, with its boresight orientated towards the rail based Rx. The Tx was placed 2m from the side of the track. The Rx height was set to 1.8m above the ground level (this was limited by railway environment regulations). Two measurement scenarios were considered: static and mobile. The static measurements were performed at spot locations every 10m along the track for Tx-Rx separation disances from 50m to 150m. For each static point scanning at the Rx was performed over 360 degrees (azimith) and over the range -15° to 15° in elevation. In the mobile measurement scenario, the Rx was pulled along the track at 1m/s from 120m to 20m from the Tx. During the mobile measurements, the azimuth and elevation angles were set to 0° at the Rx (unless stated otherwise). TABLE I. summarises the measurement parameters.

It can be seen from Fig. 1 that there are no obstructors along the measurement route between the Tx and Rx. A 3-carriage static train was located close to the start of the route. This adds some scattered components to the measured signal. On both sides of the track there are low-height building structures. There are a number of metal lighting gantries spanning the track. All measurements were performed in dry conditions.

TABLE I. MEASUREMENT PARAMETERS.

| Frequency | 60GHz |
|---|---|
| Tx height | 4m |
| Rx height | 1.8m |
| Bandwidth | 1 GHz |
| Route length | 150m |
| Transmit power at Tx | 20dBm |
| Azimuth, elevation angles at Tx | 0°, 0° |
| Antenna type | Circular horn antenna with 25dBi gain, 6.7° Half Power Beamwidth |
| Tx Polarisation | Horizontal |
| Rx Polarisation | Horizontal and vertical |

## III. MEASUREMENT RESULTS AND ANALYSIS

This section presents the results of the measurement campaign.

### A. Static Measurement Results

Fig. 3 shows the received power depends on the elevation and azimuth angles on the Rx horn. It can be seen that the maximum received power is observed with 0° elevation angle (direct Line-of-Sight (LoS) path) and 4° azimuth angle. Since the Tx was placed 2m to the side of the rail track, the Rx must be rotated approximately 4° in the azimuth plane in order to directly point towards the Tx. It is also observed that as the Tx-Rx separation distance increases, the optimum azimuth angle reduces. Some scattered paths were observed in the propagation environment. These occurred at azimuth angles significantly away from the direct LoS. Although the scattered power received at these angles, such as -50 and -160 degrees, is much lower than the LoS path, these may still be able to facilitate a viable communication link in the absence of a LoS path. It can be seen that there is around a 15dB variation (between the various curves in Fig. 3) in the maximum received power depending upon the elevation angle of the Rx antenna.

Fig. 4 shows the maximum received power at each azimuth angle depending on the received polarisation for a Tx-Rx separation distance of 50m. A 16dB difference in the maximum received power, i.e. the cross-polarisation discrimination (XPD), was observed.

Fig. 5 shows the maximum received power at each azimuth angle with respect to distance. It can be seen that the angular spread narrows with increasing distance.

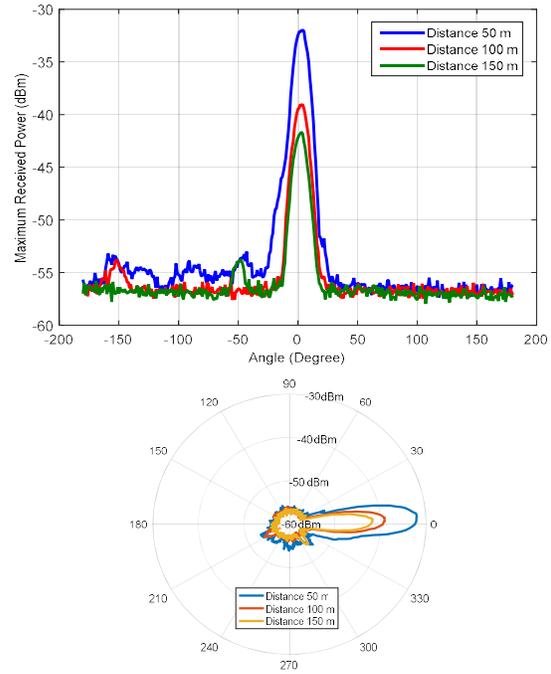

Fig. 5. Maximum received power as a function of azimuth angle and Tx-Rx separation distance.

Fig. 6 compares the maximum received power with respect to distance for co- and cross polarisations (the azimuth and elevation angles corresponding to maximum power were selected). It can be seen that over a 100m distance (from 50m to 150m) the received power drops 10dB (from -32dBm to -42dBm). We also see that the XPD reduces from 16dB to 13dB with increasing distance.

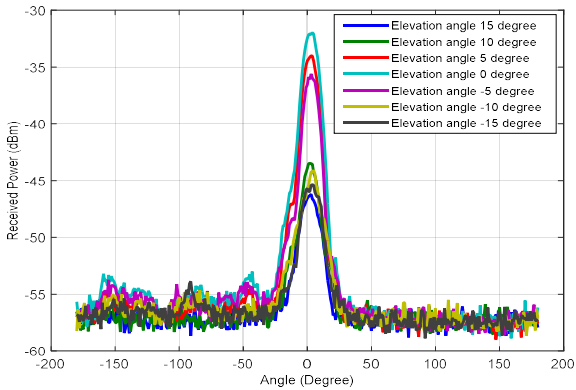

Fig. 3. Received power as a function of elevation and azimuth angle at the distance of 50m.

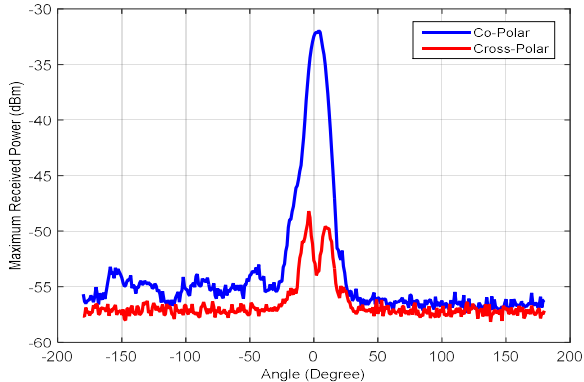

Fig. 4. Maximum received power depending on the azimuth angle at the distance of 50m for Co- and Cross-Polarisations.

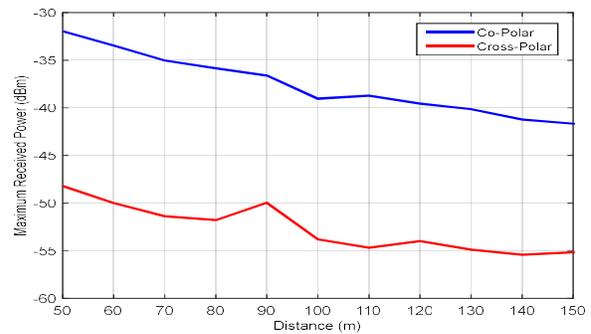

Fig. 6. Maximum received power versus distance for Co- and Cross-Polarisation.

## B. Mobile Measurement Results

In addition to the static measurements reported in the previous sub-section, dynamic measurements were also taken as the rail unit was pulled along the track. Fig. 7 shows the received power as the rail based Rx was moved from 120m to 20m from the Tx. When Fig. 7 and Fig. 6 are compared, the results can be seen to be consistent. There are only small fluctuations in Fig. 7 due to mobility. In Fig. 7, the received power decreases from 30m to 20m because of misalignment of the directional antenna to the direct path. Since the Tx and Rx heights are not equal, as the Rx approaches the Tx there is a noticeable change in elevation angle. This problem is made worse since the Tx is offset 2m from the track; hence at short separation distances there is an observable shift in the azimuth angle of the direct path. Fig. 8 compares the received power versus distance for different azimuth and elevation angles. The measurements were performed for a 15m route coving separation distances from 140m to 155m. During each measurement, the elevation and azimuth angles of the Rx antenna were fixed. It is observed that for the same azimuth angle (3°), changing the elevation angle by 3° causes up to 4dB of signal loss. Furthermore, for the same elevation angle (0°) changing the azimuth angle from 3° to 20° results in a 16dB drop in received power.

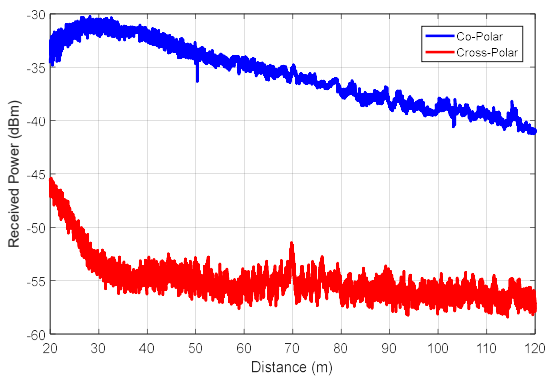

Fig. 7. Received power as the Rx moves along the track towards the Tx with the elevation and azimuth angles fixed at 0°.

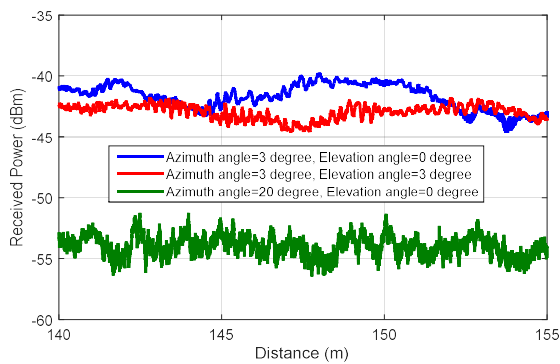

Fig. 8. Comparison of received power along the track as a function of elevation and azimuth angle.

## C. Path Loss Results

Fig. 9 illustrates the path loss results based on our static measurements. Note that for each location, the azimuth and elevation angles that provide the highest received power (i.e., the lowest path loss) were selected. This ensures that Tx and Rx directional antennas are optimally aligned. The path loss was modelled using the following path loss equation [8].

$$PL(d)[dB] = \overline{PL(d_o)} + 10n \log_{10}\left(\frac{d}{d_o}\right) + X_\sigma \quad (1)$$

where $PL(d)$ denotes the path loss at a distance $d$ from the transmitter, $\overline{PL(d_o)}$ is the path loss at a reference distance $d_o$ (in this scenario a 1m free space reference distance was chosen, hence $d_o = 1$m), $n$ is the path loss exponent and $X_\sigma$ denotes the shadow fading variance around the mean path loss. It is seen that the path loss exponent ($n = 2.04$) is very close to that of free space path loss ($n = 2$). This is to be expected since the measurement site is an open environment and the directive antennas are aligned to the LoS [9].

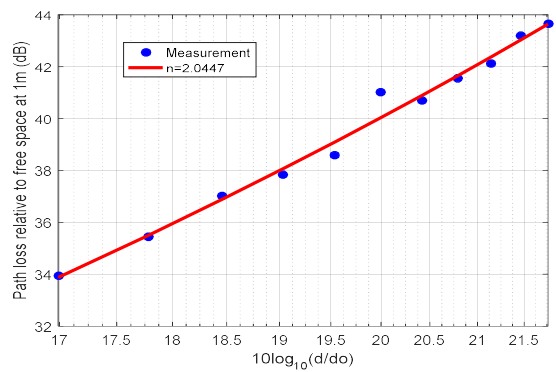

Fig. 9. Path loss versus distance.

Another important channel parameter is the RMS delay spread, which is a statistical measure of the time dispersion in the channel. The RMS delay spread is calculated directly from the measured power delay profile (PDP) [10]. Fig. 10 shows the cumulative distribution function (CDF) of the RMS delay spread. This was calculated based on the azimuth and elevation angles at the receiver that provides the highest received power for each Tx-Rx separation. For comparison, RMS delay spread values were also computed for the full range of Rx scan angles.

Only samples within 20dB of the strongest path were used when computing the RMS delay spread. RMS delay spreads for the Rx antenna orientated to the direct path were less than 1 ns. However, as shown in Fig. 10, when the boresight of the Tx and Rx antennas were not aligned, RMS delay spreads up to 22 ns were recorded. Similar results were observed in [6].

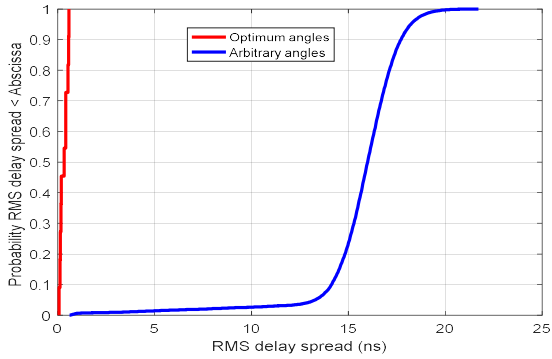

Fig. 10. CDF of RMS delay spread for optimum alignment (red) and for the complete set of scan angles (blue).

The K-factor is a key parameter for high-speed rail links [10]. Fig. 11 shows the CDF of the measured K-factor [11] when the Tx and Rx antennas were aligned. The K-factor was seen to vary between 1dB and 14.5dB. Given the very low RMS delay spreads shown in Fig. 10, the fact that 50% of channels have a K-factor less than 6dB implies multipath must exist with near identical time delay. Also, from Fig. 5, significant scatter was not seen in the azimuth plane. From this we conclude the low K-factors must be the result of ground reflections in the vertical plane between the Tx and Rx.

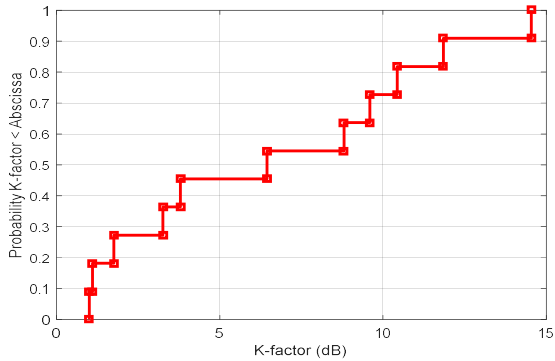

Fig. 11. CDF of K-factor for azimuth and elevation angles oriented to the direct path.

## IV. CONCLUSIONS

This paper has presented a range of mmWave channel measurements taken in a rail depot. Received data was recorded along 150m of track from a suitably placed rail side transmit unit. Channel data included path loss, angle of arrival, RMS delay spread and K-factor. High gain directional antennas were used at both ends of the link.

The measurements showed that for correct antenna alignment the path loss exponent was very close to free space path loss. RMS delay spreads ranged from 1ns to 22ns depending on antenna alignment. Even with aligned antennas, a wide range of measured K-factor values were observed. We conclude that the low K-factors are the result of ground reflections in the vertical Tx-Rx plane.

ACKNOWLEDGEMENTS

The work reported in this paper was performed as part of the RSSB (Rail Safety and Standards Board) funded MANTRA project. The authors would like to thank First Group for providing access to the St Philips Marsh rail depot.